# Phase Transition of Trapped Nuclear Exciton of Long-lived Rhodium Mössbauer States

Yao Cheng and Bing Xia

Department of Engineering Physics, Tsinghua University, Beijing 100084, P. R. China

Email: yao@tsinghua.edu.cn

We report experimental observations of the long-lived rhodium Mössbauer emissions by the time- and energy-resolved spectroscopy. The extraordinary observations manifest the open-up of photonic band gap in analogy to the superconducting gap. These observations are of potential importance for detecting gravitational waves and development of the two-photon gamma laser. Firstly, phase transitions shown by spectral evolution of characteristic emissions reveal the different aggregate exciton orderings at room temperature. Six different phases are identified by spectra profiles emitted from the color centers. Secondly, the cascade branching of the multipolar nuclear transition is discovered being the spontaneous cascade down-conversion to generate entangled γs. The macroscopic angular distribution of entangled γs from the polycrystalline sample manifests a global photon-nucleus-photon bound state across the grain boundaries. Thirdly, the γ-energy distributions depending on exciton phases reveal the photonic band gap typically on the order of several hundred eV.

In this report, we demonstrate exciton phases by the time evolution of the long-lived rhodium Mössbauer states generated by bremsstrahlung irradiation [1]. Characteristic spectral profiles of Kα, Kβ, and γ reveal several phase transitions corresponding to inversion density at room temperature.

Similar phase transitions have been observed at low temperature in [2,3]. Spectral dependence on the earth gravity manifests the 2D symmetry [3]. Phase transition depending on temperature and inversion density together enters the regime of the quasi-2D Bose-Einstein condensation of exciton, which the particular photon-nucleus-photon bound state. Phase transition of similar paths has delayed response and hysteresis to temperature variation [3] ascribed to the non-Markovian reservoir [4]. In contrast to the fixed transition patterns in [3], various paths are generally observed in this report. We thus speculate that the reported phases are mixed states of the elementary modes trapped at defects as color centers [4,5].

We discover the broad-band cascade branching channels denoted by the spontaneous cascade down-conversion (SCDC), which is analogous to the spontaneous parametric down-conversion to generate entangled γs [6]. SCDC spectrum mainly consists of two cascade photons to catalyze the E3 multipolar transition. Wavevectors, polarization, and phase of SCDC γs are matched in the nuclear Borrmann channels (NBCs) [7,8]. Reported photonic band gap (PBG) is analogous to superconducting gap depending on temperature, magnetic field and inversion density. The dependence on temperature has been reported [2,3] and its dependence on inversion density is described in this report. The dependence on magnetic field will be reported elsewhere.

Recently, Cheng *et al.* have proposed the nuclear scheme to detect gravitational waves (GWs) by utilizing the high-energy virtual states of the gravitational Stark Effect to probe the curved space-time [9]. The reported long-lived Mössbauer transitions are of potential importance for GW detection in the low frequency band. NBC is one of the focuses to achieve low-threshold γ laser [7,8]. SCDC defect mode in PBG [4,5] opens up a new approach for two-photon [10] γ laser.

Our CANBERRA detection system consists of a HPGe low-energy detector GL0510P, an optic feedback pre-amplifier 2008 BSL and a digital multichannel analyzer DSA 1000. The active detecting area is 500 mm$^2$ covering an emitting solid angle about 1.2π sr oriented in the NS direction. Bremsstrahlung irradiation and measurement set-up are the same as described in [2], besides it equipped with high-performance detector system and carried out by the fivefold intensity. The initial count rate is about 12000 cps mainly contributed by K lines in the case without filter.

Figs. 1a-1c illustrate the time evolution of spectral deformations $\tilde{S}_i(\omega,t)$ deviated from the normal profiles $\bar{S}_i(\omega)$, which are carefully calibrated. Measured spectra $S_i(\omega,t)$ are normalized by individual total counts, channel by channel in time

$$\tilde{S}_i(\omega,t) = A_i \left( \frac{S_i(\omega,t)}{\int S_i(\omega,t)d\omega} - \bar{S}_i(\omega) \right) \quad (1)$$

with factors $A_i$ chosen for each spectrum for Fig. 1. The low index $i$ stands for Kα and Kβ and γ. Phase transitions are mainly identified by satellites and hypersatellites of K lines [2]. Preliminary assignments of six phases are listed in Table 1. Of particular concern is the revival of H1 phase in Fig. 1(b), which has been observed in many other experiments. In the cavity QED [6], revival manifests the strong coupling. The reported experiments are selected to demonstrate the phase variety. To compare spectral profiles in details, we average the spectral deformations of each phase in Figs. 2a-2c. The spontaneous phase transition is shown by Fig. 2a'. Energy distributions of γ at 39.76 keV shown in Figs. 2a-2c manifest the PBG. To prove the existence of hypersatellite, Figs. 3a and 3b illustrate the stepwise reduction of coincident detection ascribed to vanishing hypersatellites after phase transitions.

SCDC γs follow the γ decay rate in time as the branching decay. Two spectra of energy > 24 keV and < 24 keV resolved by filters indicate the SCDC peak near 20 keV (see Fig. 2a and Figs. 4,5 and Fig. 3 in [3]). Total counts between 24 keV and 38 keV normalized by γ count is 0.297±0.001 in

this report, but 0.452±0.003 (Fig. 2b in [1]) and 0.446±0.003 (Fig. 2a in [1]) in [1], 0.358±0.002 in [2], 0.35±0.06 in [3] at room temperature due to different collecting solid angles. It becomes 0.59±0.03 (Fig. 1a in[3]) and 0.41±0.03 (Fig. 1b in [3]) at low temperature. Only the case of Fig. 2b in [1] is measured by the EW oriented detector, the other cases are measured by the NS oriented detector. Thus, SCDC γs emitted from the polycrystalline sample have macroscopic angular distribution. In Fig. 5, two LHS sidebands are identified at 17.2 keV and at 18.3 keV, respectively. Two sidebands are slightly different in the two filtered cases, which requires further investigation. Total SCDC counts are nearly equal to the γ counts at 40 keV. The γ escape in the blow-up figure of Fig. 4 reveals that SCDC γs are entangled [2,3].

The scheme of Fig 6a has been proposed to measure GWs [9]. Enaki and Macovei theoretically investigated the two-photon cooperative emission of dipole-forbidden transitions among N radiators [11]. They predict the collective down-conversion rate of two-photon transition enters the regime of biphoton superradiance. Frequencies $\omega_1$ and $\omega_2$ of two paired photons are in general different but peaked at $\omega_0/2$. The reported SCDC superradiance with PGB is different, which depends on dimensionality, i.e. $N^{5/3}$ for 3D isotropic NBC, $N^2$ for 2D NBC and $N^3$ for 1D anisotropic NBC [4]. Our SCDC spectral profile manifests that the scheme of two-photon cascade in Fig 6b has the major contribution. Two cascade photons ought to be the m-beam nuclear Borrmann modes (mNBMs) of similar spectral functions but opposite parity. OB wave and EB wave shown in Fig. 6b are the mNBMs of crystal symmetry such as $|3\sigma>_6$ and $|3\pi>_6$ [7] toward the (1,1,1) direction respectively.

Conventional Mössbauer nuclei emit γs rarely into 3D NBCs of a collimation angle typically on the order of $10^{-4}$ rad [7]. The NBCs of concern are particular subgroups of the general mNBMs featured by complete suppression of all elastic, inelastic resonant processes, enhanced

Lamb-Mössbauer factor and collectively enhanced nuclear coupling [7], which are particular combinations of polarized waves attached to the point-group symmetry of crystal. The situation drastically changes in SCDC branching channels, since wavevectors of varying energies freely match NBCs. The PBG of dielectric photonic crystal is fixed [5], whereas the PBGs of NBCs are dynamic in analogy to the superconducting gap. Although the averaged coupling between rhodium nuclei is weak, the SCDC coupling in NBCs is coherently enhanced by N×m for N nuclei and mNBMs [7]. SCDC localization occurs in rhodium NBCs, of which the mean free paths approach the lattice constant of 0.38 nm for N×m~30. When PBG spontaneously opens up, localized superradiance and spontaneous nuclear polarization enhance the SCDC γs intensity by $N^3$ emitting into the most significant 1D reservoir of NBCs [4] and vice versa. Though the John's $N^3$ hyperradiance and the Hannon-Trammell anomalous emission are different in many aspects, they share the same effect that all the isotropic emissions in 4π-sr vanish for sufficiently large N except in the 1D BNC. Infinitesimal impurity defects or grain boundaries in polycrystal for the anisotropic 1D case [4,5] lower the threshold of phase transition. Observed PBG is about 200 eV in this report, which requires further investigation. In analogy to the exciton of electron-hole pairs bounded by Coulomb force in semiconductors, SCDC γs are bounded by the nuclear susceptibility as the photonic potential well [5]. Moreover, in analogy to the entangled electrons in superconductors of Copper pair created by electron-phonon-electron interaction, SCDC entanglement is manifested by the photon-polariton-photon interaction, where polariton is the collective nuclear polarization in resonance with mNBMs guided in NBCs.

In conclusion, speed-up decay and exciton diffusion depending on temperature [2,3] manifest the long-lived Mössbauer Effect, though the Doppler shift [12] shall prohibit the nuclear resonance.

Our experiments show that PBG inhibits the thermal agitation on photon energy. Spontaneous open of PGB leads to Bose-Einstein condensation of exciton manifested by the photon-nucleus-photon bound states. Macroscopic angular distribution of branching SCDC γs emitted from polycrystal reveals a global photonic state. SCDC spectral profile peaked near 20 keV reveals that the biphoton SCDC transitions have major contribution. Exciton phases are identified by the emissions from exciton states pinched at defects, which are analogous to the color centers of semiconductor and the pinning center in superconductor. Nuclei coherently triggered by the long-wavelength GWs shall potentially give the important effect to detect GWs [9].

This work is supported by the NSFC grant 10675068. YC would like to thank Yi-Kang Pu for the proofreading of manuscript and J. P. Hannon for his inspiring critical comments. YC appreciates the mental support of years by Yuen-Chung Liu in the National Tsing Hua University at Hsinchu, Taiwan and the sabotage of the Tsinghua University, Beijing.

| Modes | H1 | H2 | H3 | H4 | H5 | L1 |
|---|---|---|---|---|---|---|
| Time (ks) | (a) 0-2.8, (b) 8-9.3 | (a) 5.2-11 | (b) 0-2.1 | (b) 2.1-8 | (c) 0-8.8 | (a) 2.8-5.2 |

Table 1. Preliminary assignments of exciton phases. Some of the H phases shall be mixed states of elementary modes.

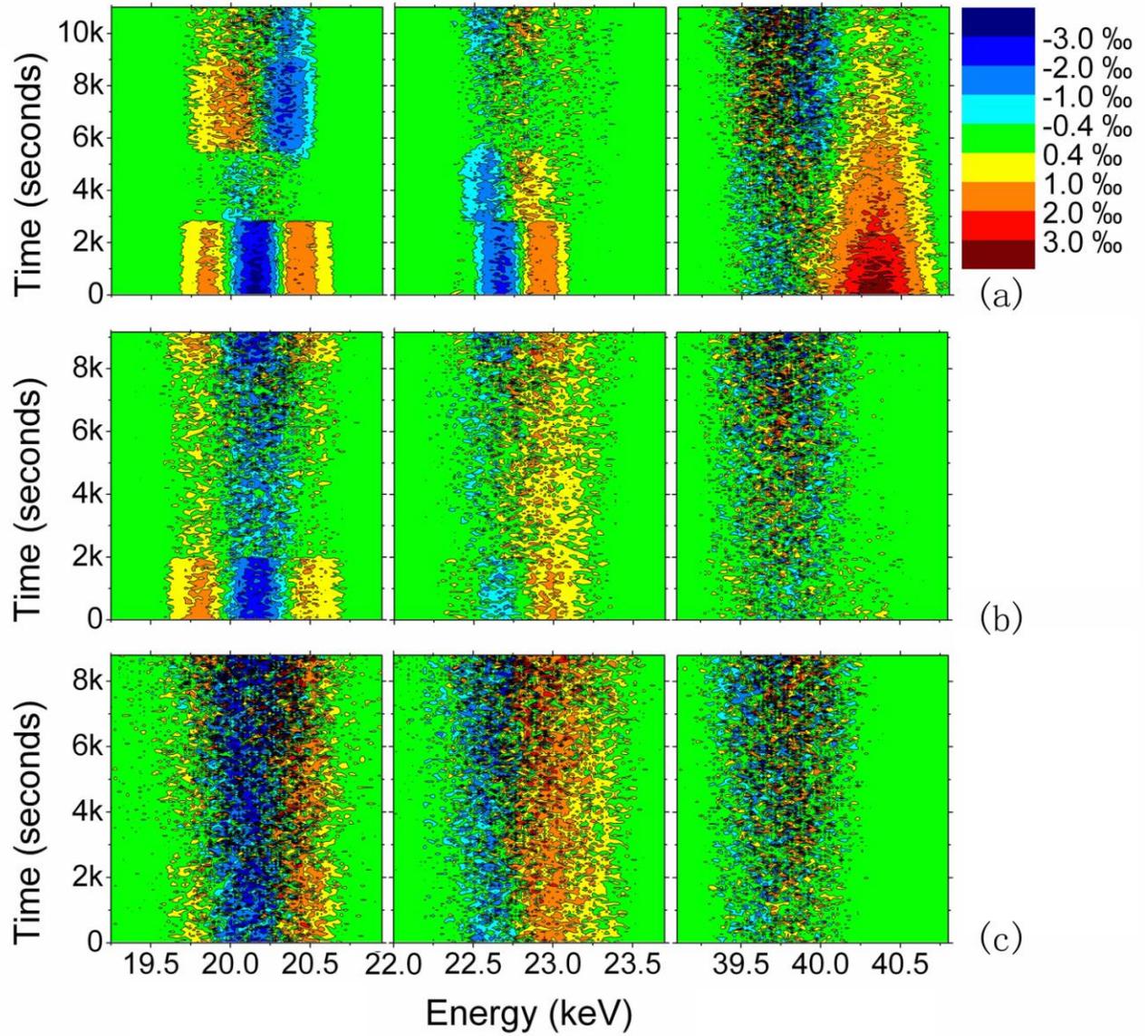

FIG. 1. (color) Time evolution of spectral deformations, i.e. Kα lines at 20 keV, Kβ lines at 23 keV, and γ at 40 keV. Total counts N(ω,t) in each band are normalized with N(Kα,t)=1, N(Kβ,t)=0.5 and N(γ,t)=0.25 for clear presentation as defined in eq. (1). Absorption sheets suppress the Kα peak pile-up located at right shoulder of the 40-keV γ peak. Filters are: (a) none; (b) 35-μm copper; (c) 25-μm tantalum. Phase assignments are listed in Table 1.

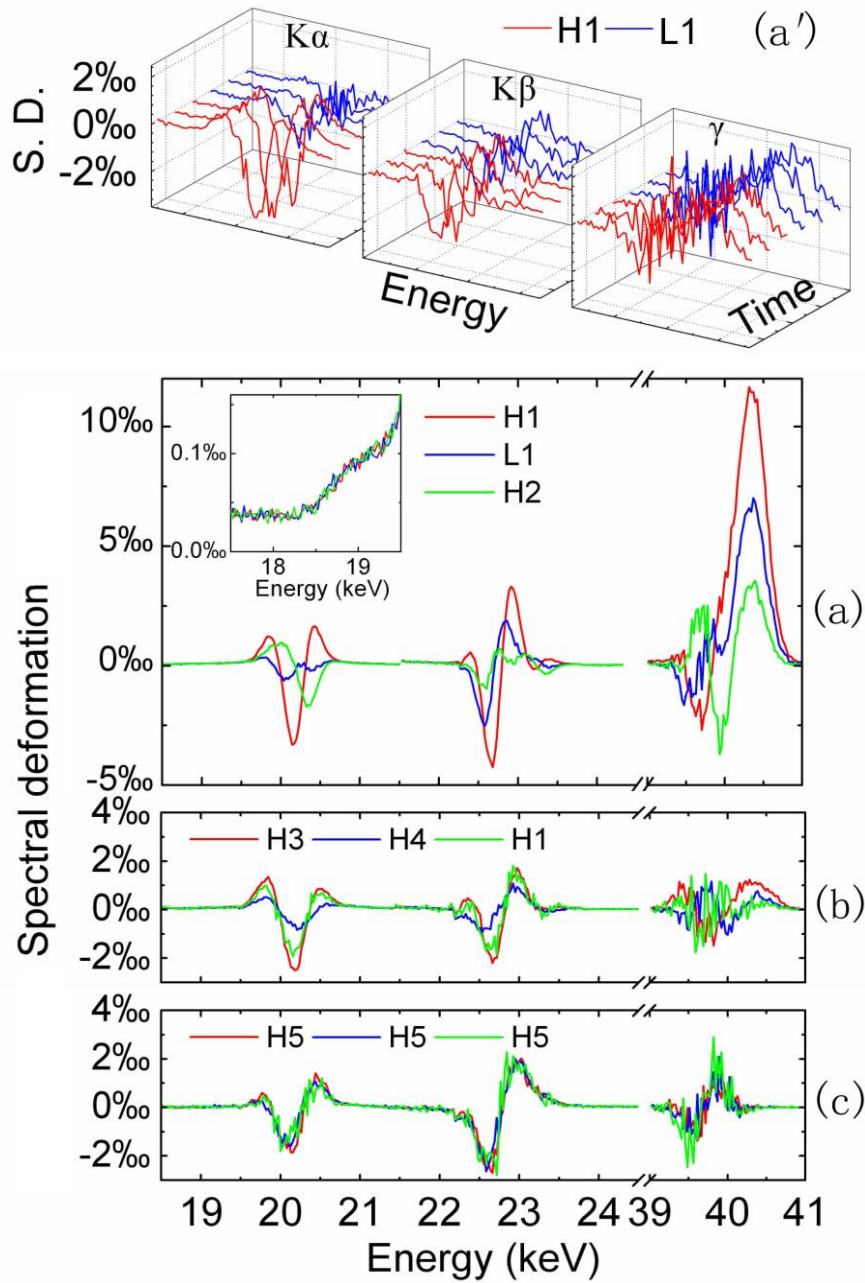

FIG. 2. (color) Averaged spectral deformations corresponding to each phase of Figs. 1(a), 1(b) and 1(c) except all N(ω,t)=1. Ordinate scales are same for three figures. (a') The original data per minute of Fig. 1a at phase transition. The ordinate scale is the same of Fig. 1a. SCDC γs near 20 keV are illustrated in the blow-up figure. Time sections of 0-2500s-6100s-8800s of (c) are arbitrarily selected.

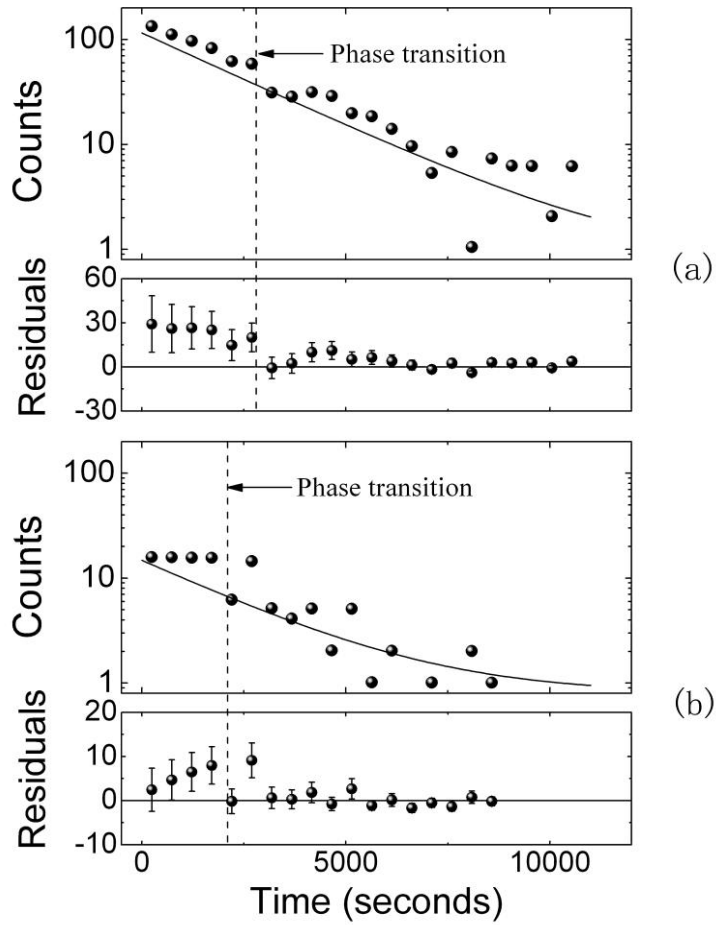

FIG. 3. Coincident counts collected between 40.8 keV and 41.3 keV corresponding to Figs. 1(a) and 1(b). Each data point is integrated for eight minutes. Residuals are calculated by removing exponential decay of 4857s/2 time constant. Phase transitions are demonstrated by the reduction of hypersatellite coincident counts: (a) from H1 to L1 at 2800s in Fig. 1(a); (b) from H3 to H4 at 2100s in Fig. 1(b).

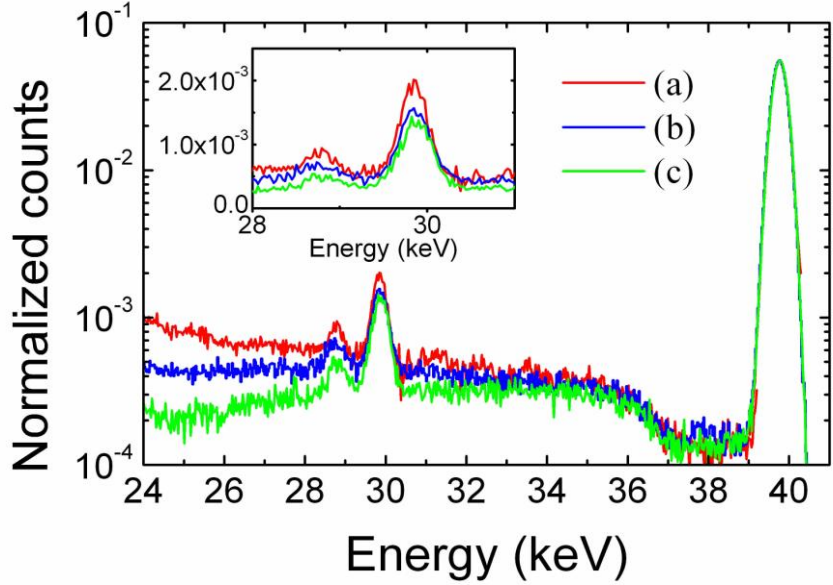

FIG. 4. (color online) SCDC spectral profile between 24 keV and 38 keV demonstrated by filters, i.e. (a) none; (b) 35-μm copper; (c) 25-μm tantalum. We have removed residual pile up by calculation. The Compton continuum (<10$^{-4}$) is negligible. SCDC counts in the order of 10$^{-3}$ are normalized by measured γ counts at 40 keV. The net counts of main escapes peak at 30 keV excluding pile-up contribution are (a) 0.023 (b) 0.019 (c) 0.018. Peaks are slightly broadened (<10%). Number 0.023 of (a) is 60% stronger than our calibration.

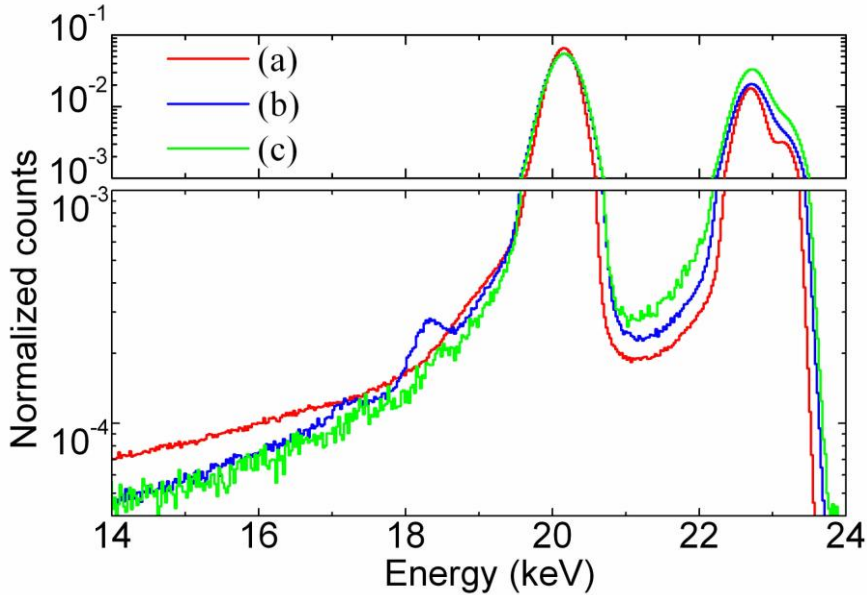

FIG. 5. (color online) SCDC spectral profile between 14 keV and 24 keV demonstrated by filters, i.e. (a) none; (b) 35-μm copper; (c) 25-μm tantalum. Spectra are normalized by the Kα. SCDC energy below Kα energy is manifested by subtracting counts with filter from counts of none-filter, whereas SCDC energy higher than Kα energy gives inversed sign. Sidebands near 17.2 keV and 18.3 keV indicate the higher transmission of entangled photons than that of individual photons. Slight peak shifts between green and blue lines are observable. Shaping times of three measurements trapezoid of (a) 5μs rise time, 0.8μs flat top (b,c) 2μs rise time, 0.6μs flat top, which leads to slightly narrower FWHM of (a) than FWHM of (b,c).

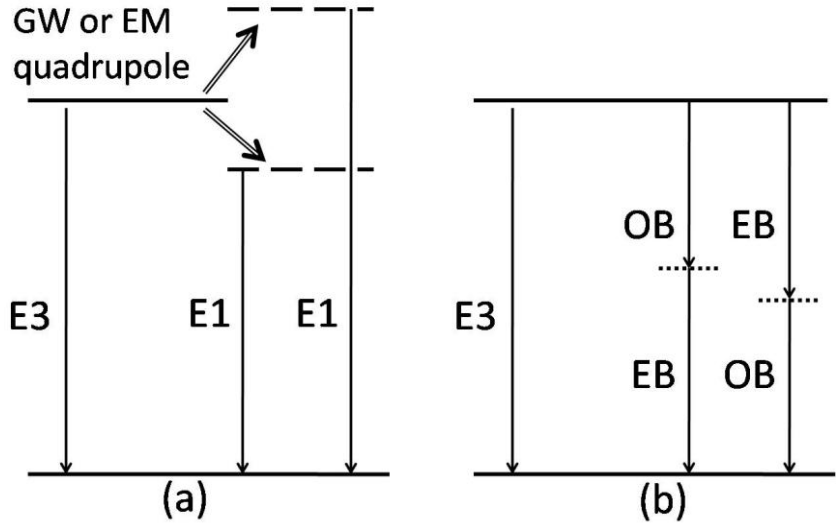

FIG. 6. Schemes to catalyze the E3 multipolar transition. Dashed lines represent the virtual states. Dot line is the floating intermediate states of SCDC transition [11]. Single-line arrows are the spontaneous emission, whereas double-line arrows are external driving field. (a) E3 multipolar transition becomes dipole transitions induced by gravitational Stark effect [9] or by thermal electro-magnetic quadrupole field [11]. (b) SCDC speed-up scheme. EB and OB are mNBMs of even parity and odd parity respectively.